# An Adaptive Power Division Strategy for Nonlinear Components in Rectification

Zhongqi He, Liping Yan, *Senior Member, IEEE*, and Changjun Liu, *Senior Member, IEEE*

***Abstract*—This letter presents a novel adaptive power division strategy, which uses two rectifying diodes with nonlinear impedance characteristics that are configured in parallel to function optimally at their individual power levels. Through strategic adjustment of the input admittance, the conductance of the low-power diode decreases progressively with increasing power, while the conductance of the high-power diode increases correspondingly. This conductance-based power allocation method ensures that the power is rectified consistently by the most appropriate diode, regardless of the power level, and thus enables efficient rectification across an extended range. This letter presents a rectifier typology to substantiate the proposed strategy. Experimental results confirm the efficiency of the adaptive power division strategy, with the rectifier showing efficiency in excess of 60% from 5 dBm to 29.5 dBm, giving a power dynamic range of 24.5 dB.**

## I. Introduction

In wireless power transmission (WPT) systems, such as powering unmanned aerial vehicles, the challenge of fluctuating received power levels persists due to the mobility of transmitters/receivers, influenced by transmission distance, antenna orientation, and polarization. To tackle this, it is essential to maintain high microwave-to-dc conversion efficiency across different power levels.

Rectifiers use diodes to convert microwave power into dc power [1][2]. However, diodes are inherently nonlinear and exhibit variable input impedance characteristics that are dependent upon the received power level. These fluctuations cause impedance mismatches that reduce the rectification efficiency.

Research efforts have been made previously to expand the operating power ranges of rectifiers. In [3], a rectifier with an impedance compression network achieved a power dynamic range of 17.2 dB. A dual-band resistance compression network was presented in [4]. Ref. [5] describes the use of the output dc voltage to adjust the capacitance of a parallel varactor diode, thus compensating for the impedance shifts in the rectifying diode caused by input power variations. Balun structures have been implemented to stabilize input impedance variations [6], while diode-cooperative structures have been used to improve the power range [7]. Recent studies have also explored wide power dynamic range rectifiers that operate based on a power-adaptive division approach [8]-[10].

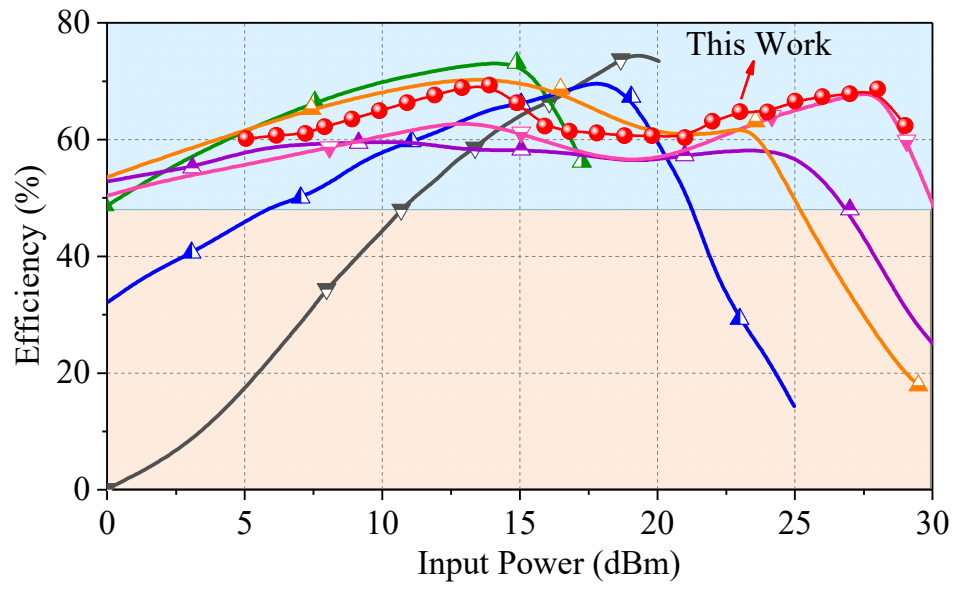

Fig. 1. Measured efficiencies of 2.4 GHz-band rectifiers with wide power ranges from recent publications. The detailed data are listed in Table I.

TABLE I
PARAMETERS OF RECTIFIERS IN FIG. 1

| Symbol | Year | Freq. (GHz) | Diode | Publication Title | Ref. |
|---|---|---|---|---|---|
| (gray ▽) | 2017 | 2.34 | HSMS 286×2 | IEEE TMTT | [6] |
| (green △) | 2019 | 2.45 | HSMS 286×2 | IEEE Access | [3] |
| (blue △) | 2019 | 2.4 | HSMS 286×4 | IEEE TCAS I | [4] |
| (purple △) | 2019 | 2.4 | HSMS 282×1<br>SMS 7630×1 | IEEE TMTT | [7] |
| (pink ▽) | 2021 | 2.45 | HSMS 282×2<br>HSMS 286×1 | IEEE TPE | [8] |
| (orange △) | 2024 | 2.4 | HSMS 282×1<br>HSMS 286×1 | IEEE MWTL | [10] |

Couplers [11], and power dividers [12] are also used to design wide power range rectifiers. The fixed power division ratios of couplers and power dividers prevent the power from being rectified by the most suitable rectifier and thus limit the dynamic power range. In [13], the corresponding switch is activated based on the input power to select the appropriate rectifier branch. However, this design requires an additional power supply.

In the design presented in this work, two rectifiers are connected in parallel. Two transmission lines are used to adjust the input conductance values of these rectifiers. Ultimately, the input conductance of the low-power rectifier decreases monotonically. In contrast, the input conductance of the high-power rectifier increases monotonically, and adaptive power distribution between the parallel rectifiers is achieved. This method enhances the rectification efficiency significantly across a wide power range, as shown in Fig. 1.

Manuscript received ——, ——.

This work was supported in part by the Natural Science Foundation of China (NSFC) under Grant U22A2015, the National Funded Postdoctoral Researcher Program of China under Grant GZC20231766, and the Sichuan Science and Technology Program under Grant 2024YFHZ0282. (Corresponding author: *Changjun Liu*.)

The authors are with the School of Electronics and Information Engineering, Sichuan University, Chengdu 610064, China (e-mail: cjliu@ieee.org).

Zhongqi He and Changjun Liu are also with the Yibin Industrial Technology Research Institute, Sichuan University, Yibin 644005, China.

## II. Principle of Unequal Power Division Strategy

Admittance, as the reciprocal of the impedance, is a critical parameter in this context. We have used electromagnetic simulation software to model the relationship between impedance, admittance, and the input power directly for the HSMS 282 and HSMS 286 diodes at 2.4 GHz.

The pronounced nonlinearity of the impedance and admittance characteristics with power variations, shown in Fig. 2, represents a remarkable challenge when designing rectifiers capable of operation with a wide power dynamic range.

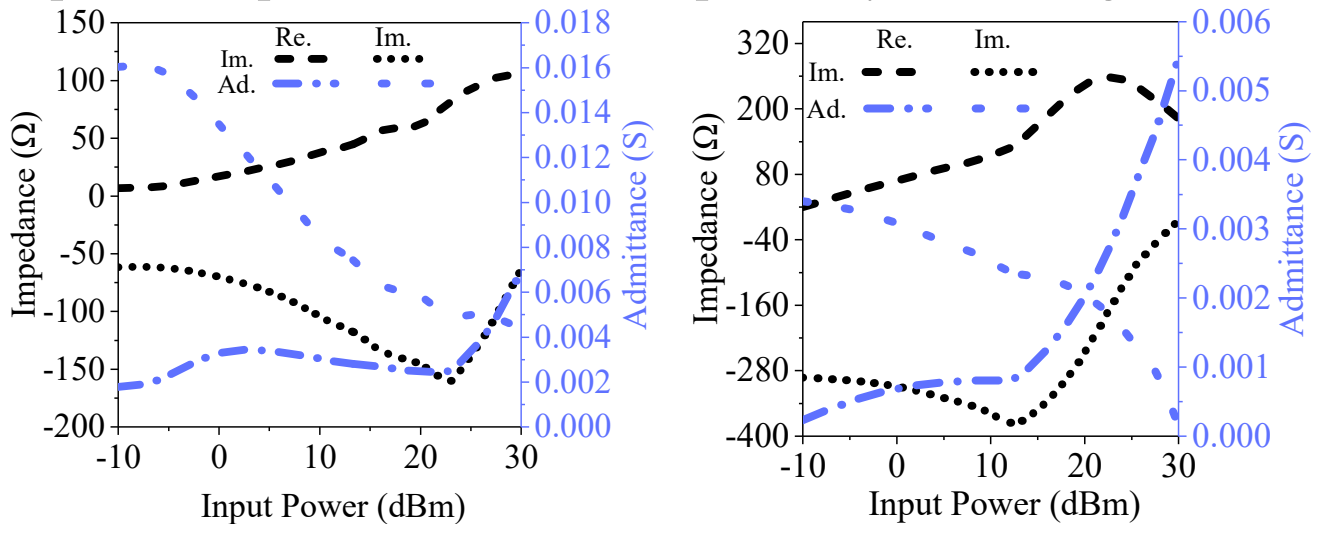

Fig. 2. (a)Simulated impedance and admittance vs. power for the HSMS282 diode at 2.4 GHz with a load of 600 Ω, and (b)for HSMS286 diode at 2.4 GHz with a load of 1200 Ω.

The proposed power division structure is depicted schematically in Fig. 3. The characteristic admittance and length of transmission line $TL_1$ are denoted by $Y_{01}$ and $l_1$, and $Y_{02}$ and $l_2$ are the characteristic admittance and length of transmission line $TL_2$, respectively.

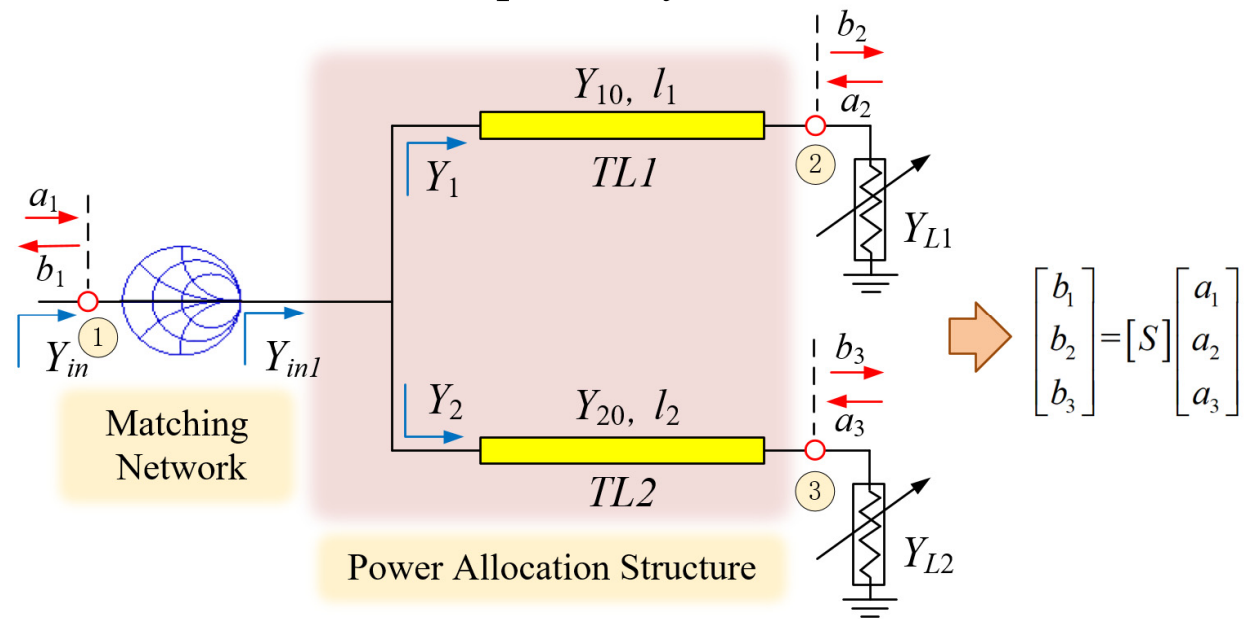

Fig. 3. Schematic showing the proposed power division strategy.

The $S$-parameters of the three-port network are calculated as follows:

$$[S]=\begin{bmatrix} S_{11} & S_{12} & S_{13} \\ S_{21} & S_{22} & S_{23} \\ S_{31} & S_{32} & S_{33} \end{bmatrix} \tag{1}$$

and

$$\frac{|S_{21}|^2}{|S_{31}|^2}=\frac{\mathrm{Re}(Y_1)}{\mathrm{Re}(Y_2)}=\frac{Y_{01}\dfrac{Y_{01}Y_{L1}\left(1+\tan^2\beta l_1\right)}{Y_{01}^2+Y_{L1}^2\tan^2\beta l_1}}{Y_{02}\dfrac{Y_{02}Y_{L2}\left(1+\tan^2\beta l_2\right)}{Y_{02}^2+Y_{L2}^2\tan^2\beta l_2}} \tag{2}$$

Connecting two identical rectifying diodes in parallel necessitates adaptive power distribution across varying power levels to extend the power dynamic range of the rectifiers. For example, at lower power levels, one diode should dominate power absorption, while the other diode should accept the majority of the power at higher power levels.

As illustrated in Fig. 3, the two branches of the power division structure are similar; therefore, analysis of a single branch alone is sufficient. We thus focus on the characteristics of the upper branch.

When the load admittance $Y_{L1}$ is that of a rectifying diode, and assuming that $\mathrm{Re}(Y_{L1})$ is approximately 0.004 S (as shown in Fig. 2), the relationship between the input admittance $Y_1$, the transmission line characteristic admittance $Y_{01}$, and the line length $l_1$ is as shown in Fig. 4(a). $Y_1$ is on a constant reflection coefficient circle on the Smith chart.

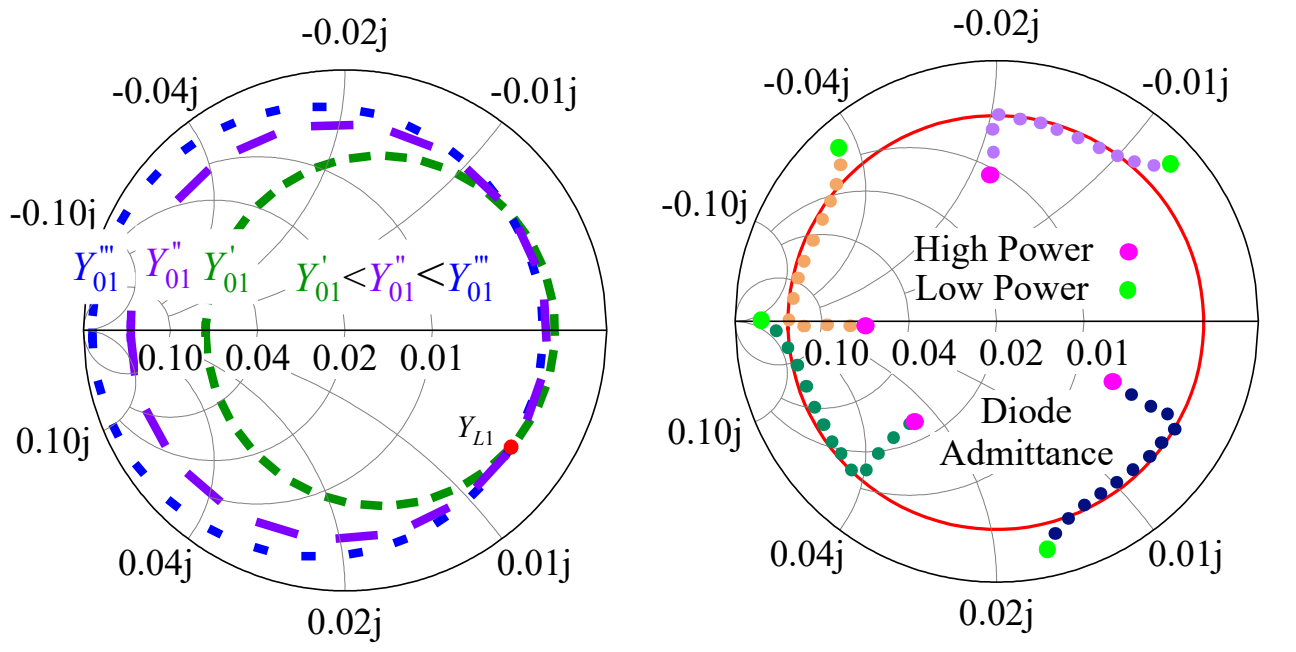

Fig. 4. (a)Relationships between $Y_1$, $Y_{01}$, and $l_1$. (b) Relationship between $Y_1$ and $l_1$ when $Y_{L1}$ is the diode admittance.

When $Y_{01}$ is low, variation in $Y_1$ along with $l_1$ results in a relatively small change in $\mathrm{Re}(Y_1)$, as illustrated by the green line in Fig. 4(a). In contrast, when $Y_{01}$ is high, $\mathrm{Re}(Y_1)$ shows a larger variation range, as shown by the blue line.

The power division of the parallel diode branches is directly proportional to the conductance of the input admittance, i.e., $\mathrm{Re}(Y_1)$. The desired power division strategy can thus be achieved by tuning $\mathrm{Re}(Y_1)$. To investigate the input admittance further with respect to the diode admittance variation, we set $Y_{L1}$ to be the admittance of the diode from low power to high power. The input admittance $Y_1$ is dependent on the transmission line length $l_1$, as shown in Fig. 4(b).

Fig. 4(b) shows that the input admittance $Y_1$ can change its variation range with different values of transmission line length $l_1$. Furthermore, the relationship between the applied power and the admittance may be inverted.

We pay increased attention to the conductance of the diode. At low power levels, the conductance of the diode remains relatively small. In contrast, the conductance increases at high power levels, as indicated by the dark blue dashed line in Fig. 4(b). However, the conductance shows only a minimal variation range between the high and low power levels, remaining at approximately 0.004 S.

We analyze the input admittance here at three typical transmission line lengths.

**Case 1:** When $l_1$ is approximately λ/8, the admittance of the diode rotates by about π/4. The input conductance becomes high under low power conditions and low under high power conditions.

**Case 2:** When $l_1$ approaches λ/4, the admittance of the diode is then transformed from inductive to capacitive under low power conditions, and the conductance decreases. Under high power conditions, the conductance changes from low to high. Additionally, the conductance values under the high and low power conditions still exhibit a significant variation range.

**Case 3:** When $l_1$ approaches λ/3, the admittance of the diode then exhibits capacitive susceptance throughout the entire power range. The conductance of the diode remains low at low power and high at high power.

Based on the analysis above, we can select $TL_1$ to have different lengths to obtain the desired conductance, thereby allowing the expected power distribution to be realized.

The power division strategy is designed to be in three states:

(1) Low power level

When the input power is low and Re($Y_1$) >> Re($Y_2$), the relationship between the reflected powers from port 2 and port 3 is then:

$$\frac{|b_2|^2}{|b_3|^2} \to \infty \tag{3}$$

This means that $Y_{L1}$ can acquire most of the input power under low-power conditions.

(2) Medium power level

When the input power is at a medium level and Re($Y_1$) = Re($Y_2$), then the relationship between the reflected powers of port 2 and port 3 can be obtained as follows:

$$\frac{|b_2|^2}{|b_3|^2} \approx 1 \tag{4}$$

This means that $Y_{L1}$ and $Y_{L2}$ can acquire half the input power each under medium power conditions.

(3) High power level

When the input power is high and Re($Y_1$) << Re($Y_2$), the relationship between the reflected powers of port 2 and port 3 can be determined as follows:

$$\frac{|b_2|^2}{|b_3|^2} \to 0 \tag{5}$$

This means that $Y_{L2}$ can acquire most of the input power under high power conditions.

Therefore, by varying the length of the transmission line connected in series at the front of the parallel diodes, the conductance values of the diodes at the different powers can be changed and unequal power division can ultimately be achieved. It is thus possible to operate the two diodes at various power levels and finally expand the power dynamic range of the rectifier.

## III. Design, Simulation, and Measurement of the Proposed Rectifier

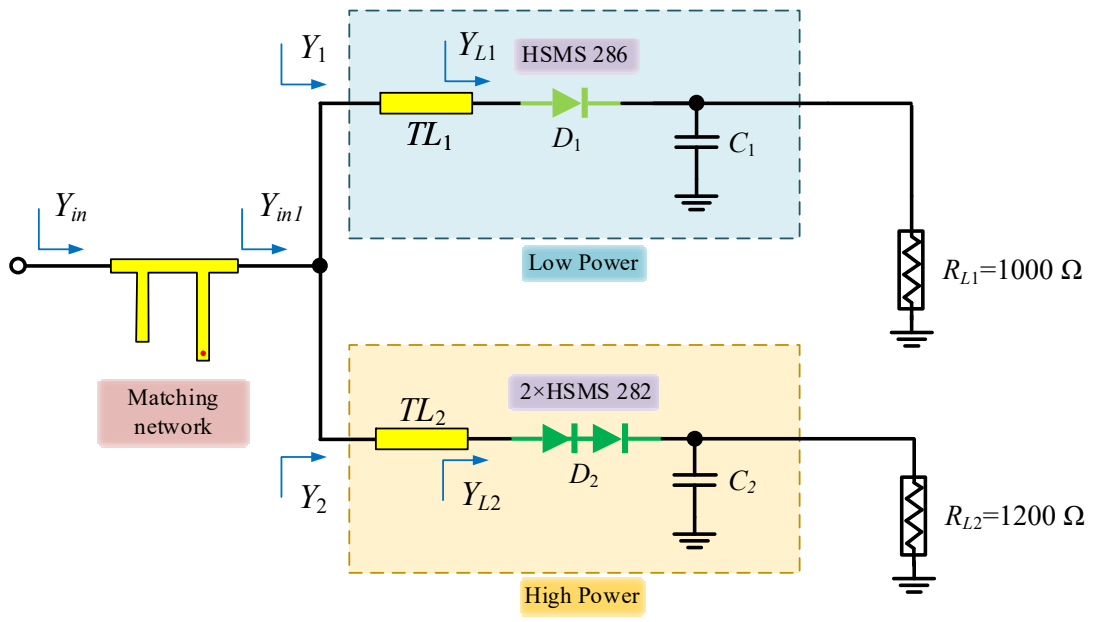

Fig. 5. Schematic of the proposed rectifier.

Different combinations of diodes can be used to expand the power dynamic range of the rectifier further. In the proposed rectifier, the low-power diode is an HSMS 286 diode, while the high-power diodes, consist of two series-connected HSMS 282 diodes. The series rectification structure means that no dc-blocking capacitors are required to be inserted before the diodes, and this contributes to a further reduction in the insertion loss of the rectifier. The proposed rectifier is shown schematically in Fig. 5.

### *A. Design and Simulation*

The input admittances of diodes $D_1$ and $D_2$ were simulated, with results as shown in Fig. 6(a). The conductance characteristics of both diodes show minimal variations across the entire power range. Subsequently, the proposed power distribution strategy is used to design the rectifier. The characteristic admittance values and the electrical lengths of the two series-connected transmission lines designated $TL_1$ and $TL_2$, are first optimized in the power divider network. The simulation results for the input admittance values $Y_1$ and $Y_2$ of the diodes after transformation through the transmission lines are presented in Fig. 6(a), and these results indicate that an enhanced diode conductance variation range is realized.

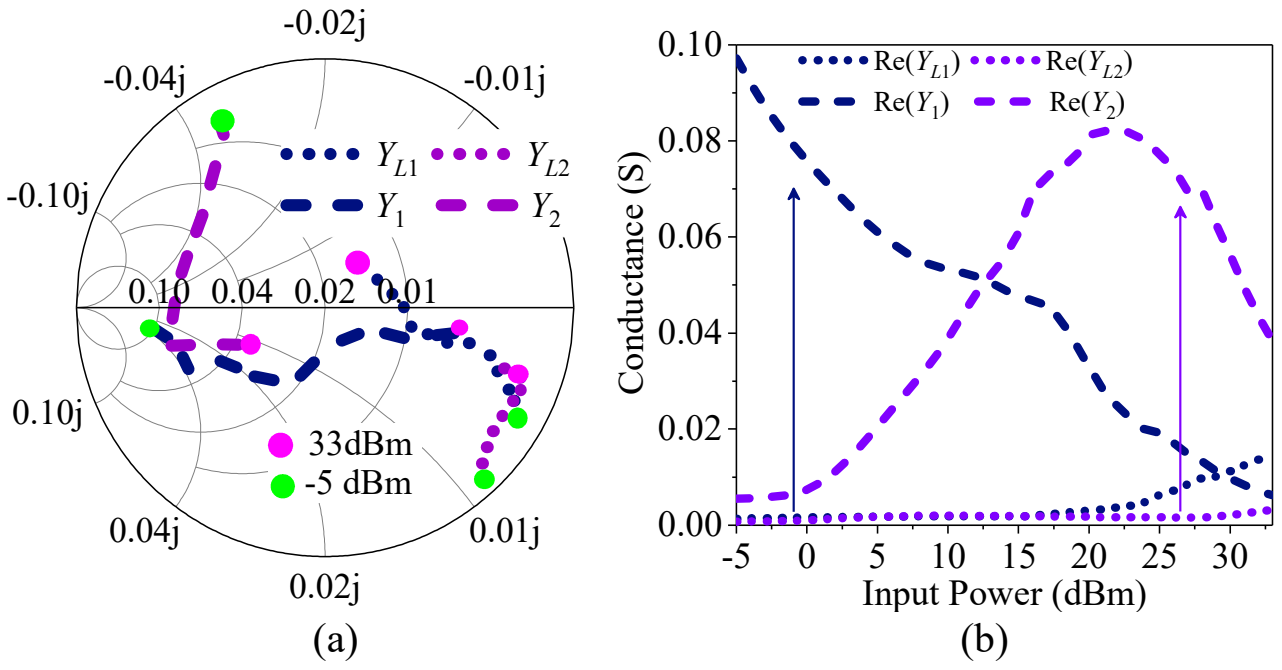

Fig. 6. (a)Simulated $Y_{L1}$, $Y_{L2}$, $Y_1$, and $Y_2$ of the proposed rectifier on a Smith chart at different powers. (b) Simulated Re($Y_{L1}$), Re($Y_{L2}$), Re($Y_1$), and Re($Y_2$).

To investigate the conductance variations further, Fig. 6(b), depicts the simulated Re($Y_{L1}$), Re($Y_{L2}$), Re($Y_1$), and Re($Y_2$) values of the proposed rectifier at various powers. The figure shows that diodes $D_1$ and $D_2$ experience a reduction and an increase in their conductance at low- and high-power levels, respectively, thus enabling the implementation of the power distribution strategy.

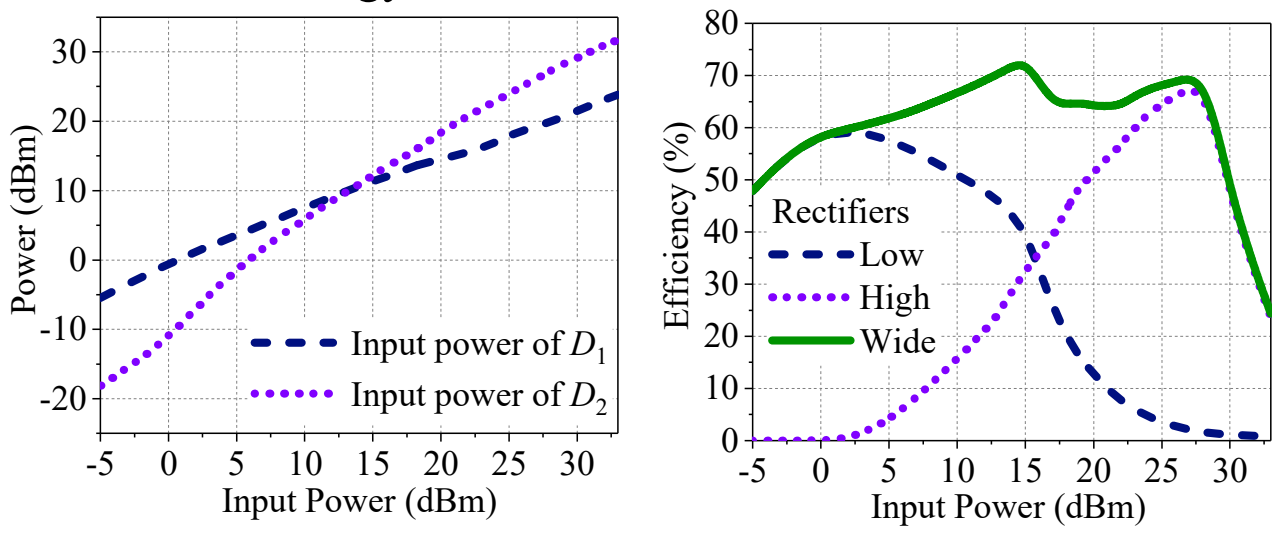

Fig. 7. (a)Power division of the proposed rectifier. (b)Simulated efficiency characteristics of the proposed rectifier and the efficiencies of each diode at different input powers.

Fig. 7(a) shows the simulated input power characteristics of diodes $D_1$ and $D_2$. Low-power diode $D_1$ receives most of the input power at low powers. Conversely, the high-power diode $D_2$ receives most of the power at high powers. Therefore, this rectifier design can achieve an adaptive power distribution strategy.

By continuing with the simulation, the efficiency of the proposed rectifier is also analyzed. As depicted in Fig. 7(b), at low power levels, diode $D_1$ operates predominantly. At medium power levels, diodes $D_1$ and $D_2$ work together in tandem. At high power levels, the rectifier switches to diode $D_2$ operation.

Additionally, we simulated the impact of temperature on the rectifier. When the temperature rises to 100°C, the maximum efficiency of the rectifier decreases by 7% compared to at -25°C. Furthermore, in practical applications, it is necessary to consider the heat dissipation of high-power diodes. By enhancing the thermal design, the high performance of the diodes under high power can be effectively maintained.

### *B. Measurement*

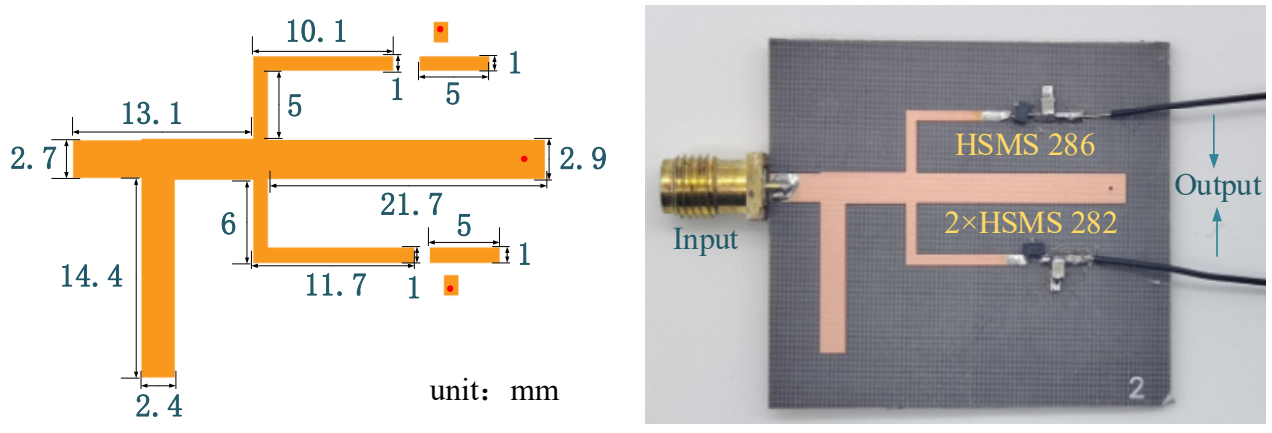

(a) (b)

Fig. 8. Proposed rectifier. (a) Layout. (b) Photograph.

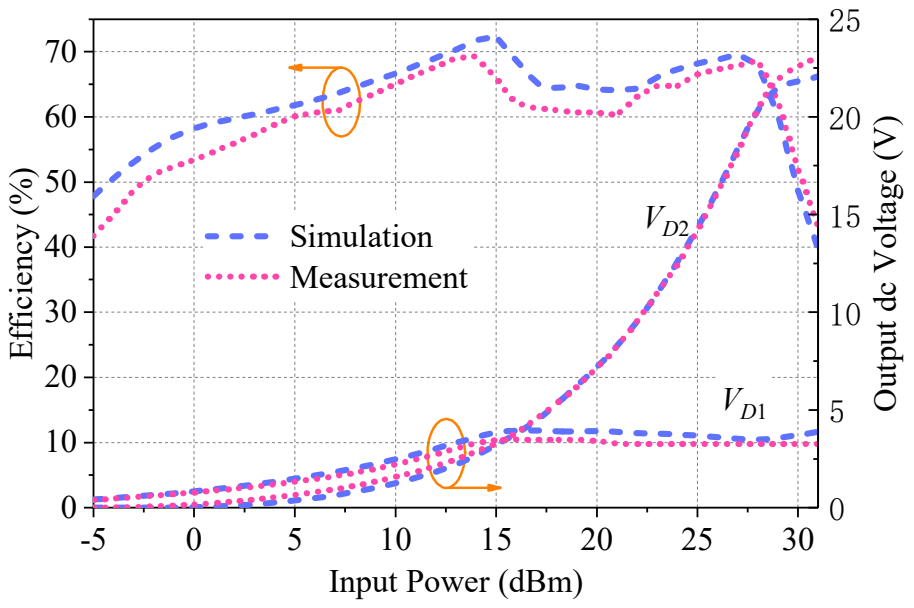

Fig. 9. Simulated and measured RF-dc conversion efficiency and output dc voltage characteristics of the proposed rectifier at various input powers.

TABLE II
COMPARISON WITH PRIOR WIDE POWER RANGE RECTIFIERS

| **Ref.** | **Year** | **Freq. (GHz)** | **Power range for efficiency>50%** | **Power range for efficiency>60%** | **Dimension** |
|---|---|---|---|---|---|
| [6] | 2017 | 2.34 | >13 dB | >9 dB | 95 mm×62 mm |
| [5] | 2018 | 2.4 | 2.5 dBm–25.5 dBm (23 dB) | 7 dBm–25 dBm (18 dB) | 30 mm×19 mm |
| [3] | 2019 | 2.45 | 0.5 dBm–18.5 dBm (18 dB) | 4 dBm–17 dBm (13 dB) | 43 mm×35 mm |
| [9] | 2021 | 2.45 | 9.9 dBm–30.8 dBm (20.9 dB) | 15 dBm–26 dBm (11 dB) | 26 mm×20 mm |
| [10] | 2024 | 2.4 | -1.6 dBm–24.4 dBm (26 dB) | 4 dBm–24 dBm (20 dB) | 40 mm×35 mm |
| **This work** | **2024** | **2.43** | **-2.5 dBm– 30.1 dBm (32.6 dB)** | **5 dBm–29.5 dBm (24.5 dB)** | **39 mm×35 mm** |

The layout and a photograph of the proposed rectifier are shown in Fig. 8. The test results are presented in Fig. 9. When the input power ranges from −2.5 to 30.1 dBm (32.6 dB), the rectification efficiency exceeds 50% consistently. Moreover, the rectification efficiency remains above 60% over the entire input power range from 5 to 29.5 dBm, thus indicating a high-efficiency power dynamic range of 24.5 dB.

Table II compares the proposed rectifiers with several prior works. Generally, power dynamic rectifiers focus on power ranges with efficiencies greater than 50%. Because of the use of the high-performance adaptive power distribution structure in this design, wide-range rectification can be achieved with higher efficiency. The power range of the rectifier, as proposed in this letter demonstrates efficiency of more than 60%, which is the widest range.

## IV. CONCLUSION

A high-efficiency RF rectifier arranged in an adaptive power distribution structure with an extended input power range has been proposed and designed using two rectifying diodes. The adaptive power division strategy proposed here can be also applied to design other RF circuits in which unequal power division of different branches is required.